\documentclass{ceab}   %% loading ceab.cls creates CEAB style
\usepackage{amsmath}
\usepackage{upgreek}
\usepackage{epsfig}     %\  Used to include figures
\usepackage{graphicx}   %/
\usepackage{url}
\usepackage{multirow}
\usepackage{multicol}
\usepackage{natbib}     % for producing References section 
\usepackage[T1]{fontenc} % for producing Polish vowels 
\usepackage{babel}       % for different languages
    
\def\runningtitle{Owen et al.: Cosmic ray neutrons in magnetized astrophysical structures}   
\def\articlenumber{7}
\def\ppages{1--4}

\usepackage{tikz}

\newcommand{\myarrow}{\tikz\draw[black,->] (1,4.5ex) -- ++(0,-2.05ex) -- +(4.5ex,0); \;}

\begin{document}

\title{Cosmic ray neutrons in magnetized astrophysical structures}

\author{Ellis R. Owen$^{1,2}$\thanks{Speaker}, Yoshiyuki Inoue$^{1,3,4}$, \\ 
Tatsuki Fujiwara$^{1}$, Qin Han$^{5}$ and Kinwah Wu$^{5,3}$ 
\vspace{2mm}\\
\it $^{1}$Theoretical Astrophysics, Department of Earth and Space Science, \\ 
\it Graduate School of Science, The University of Osaka,\\ 
\it Toyonaka 560-0043, Osaka, Japan  \\
\it $^{2}$Astrophysical Big Bang Laboratory (ABBL), RIKEN Pioneering Research Institute, \\
\it Wak\={o}, Saitama, 351-0198 Japan \\ 
\it $^{3}$Kavli Institute for the Physics and Mathematics of the Universe (WPI), \\
\it UTIAS, The University of Tokyo, Kashiwa, Chiba 277-8583, Japan \\
\it $^{4}$Interdisciplinary Theoretical \& Mathematical Science Program (iTHEMS), \\ 
\it RIKEN, 2-1 Hirosawa, Saitama 351-0198, Japan \\
\it $^{5}$Mullard Space Science Laboratory, University College London, \\
\it Holmbury St.~Mary, Surrey RH5 6NT, UK\\
}

\maketitle

\vspace{-1cm}
\begin{abstract}
Cosmic rays are often modeled as charged particles. This allows their non-ballistic propagation in magnetized
structures to be captured. In certain situations, a neutral cosmic ray component can arise. For example, cosmic ray neutrons are produced in considerable numbers through hadronic pp
and p$\gamma$ interactions. At ultrahigh energies, the decay timescales of these neutrons is dilated, allowing them to traverse distances on the scale of galactic and cosmological structures. Unlike charged cosmic rays, neutrons are not deflected by magnetic fields. They propagate ballistically at the speed of light in straight lines. The presence of a neutral baryonic cosmic ray component formed in galaxies, clusters and cosmological filaments can 
facilitate the escape and leakage of cosmic rays from magnetic structures that would otherwise confine them. We show that, by allowing confinement breaking, the formation of 
cosmic-ray neutrons by high-energy hadronic interactions in 
large scale astrophysical structures can modify the exchange of ultra high-energy particles across magnetic interfaces between galaxies, clusters, cosmological filaments and voids. 
\end{abstract}

\keywords{Cosmic rays - cosmic web (filaments) - galaxies - galaxy clusters}

\sloppy
%
%====================================================== 
%======================================================
\section{Introduction}

Cosmic rays are an inhomogeneous collection of energetic particles, including hadrons and leptons. In many studies, cosmic rays are considered to be charged, but they also consist of neutral particles. 
This neutral component can be substantial in certain situations. For instance, neutrons are abundant in atmospheric cosmic-ray showers 
\citep[e.g.][]{Schimassek2024arXiv}, 
and most of the cosmic-rays baryons 
detected at sea level on Earth are actually neutrons 
\citep[see][]{Ziegler1998IBMJ, Sato2015PLoSO}. 

Free neutrons have a lifetime of $879.6 \pm 0.8~{\rm s}$
\citep{PDG2020PTEP} in their rest frame. 
The lifetimes of energetic neutrons are significantly longer 
due to relativistic time dilation. 
High-energy neutrons 
can therefore propagate uninterrupted
over large distances 
before they decay. 
A 100 PeV neutron can easily cross galaxy-scale structures, 
and a $10~{\rm EeV}$ neutron 
is able to traverse a galaxy cluster or super-cluster.

%
%======================================================
%======================================================
\section{Cosmic-ray baryons in astrophysical environments}

%
%======================================================
\subsection{Hadronic Interactions} 

Ultra high-energy (UHE) cosmic-ray particles are generally produced by acceleration in violent 
astrophysical environments. 
This requires that the particles being accelerated are charged, and that a magnetic field is present at the acceleration site \citep{Fermi1949PhRv,Bell1978MNRAS}.  
Charged cosmic-ray particles that break  
   magnetic confinement 
   and leave their acceleration site   
   can interact with ambient matter 
   via a pp pion-production process if their energies are above a threshold of about  
  $280~{\rm MeV}$\footnote{This is the energy required for the production of a neutral pion through the channel pp$\rightarrow$pp$\pi^0$}. 
The dominant channels in the pp process are: 
\begin{align}
\label{eq:pp_interaction} 
{\rm p} + {\rm p} \longrightarrow 
	\begin{cases} 
	\; 	{\rm p}\;\!  \Delta^{+~} \;\! \ \longrightarrow  \begin{cases} 
				{\rm p}\;\! {\rm p}\;\! \pi^{0} \;\! \xi_{0}(\pi^{0})\;\! \xi_{\pm}(\pi^{+} \pi^{-}) \\[0.5ex] 
				{\rm p}\;\! {\rm p}\;\!  \pi^{+}  \pi^{-}\;\!  \xi_{0}(\pi^{0})\;\! \xi_{\pm}(\pi^{+} \pi^{-}) \\[0.5ex] 
				{\rm p}\;\! {\rm n}\;\!  \pi^{+}\;\!  \xi_{0}(\pi^{0})\;\! \xi_{\pm}(\pi^{+} \pi^{-})\\[0.5ex] 
			\end{cases} \\ 
	\; 	{\rm n}\;\! \Delta^{++} \longrightarrow  
            \begin{cases}
				{\rm n}\;\! {\rm p}\;\! \pi^{+}\;\! \xi_{0}(\pi^{0})\;\! \xi_{\pm}(\pi^{+} \pi^{-}) \\[0.5ex]
				{\rm n}\;\! {\rm n}\;\! \pi^+ \pi^{+} \xi_{0}(\pi^{0})\;\! \xi_{\pm}(\pi^{+} \pi^{-})\\[0.5ex]
			\end{cases} \\
	\end{cases} \ .  
\end{align}%  
Here, $\xi_0$ and $\xi_{\pm}$ are the multiplicities for neutral and charged pions, respectively, and  
 the $\Delta^+$ and $\Delta^{++}$ baryons are resonances. 
Cosmic-ray neutrons 
  are produced 
  both directly as secondary baryons 
  and through resonance decays. 
Energetic cosmic-ray baryons 
 can also interact with 
 radiation fields via p$\gamma$ (or n$\gamma$) interactions. 
For p$\gamma$ processes, 
 the dominant channels are resonant single-pion production, 
 direct single-pion production, and multiple-pion production \citep[e.g.][]{Mucke1999PASA}.  
Resonant single-pion production occurs through the formation of $\Delta^{+}$ particles, which decay through two major channels. Both of these channels produce charged and neutral pions, secondary protons and neutrons: 
\begin{align}%
\label{eq:pg_int}%
 {\rm p}+  \gamma \longrightarrow 
   %\Delta^{+}&  \longrightarrow    
	\begin{cases}% 
	  \;  {\rm p} \pi^0 \longrightarrow {\rm p}\;  2\gamma				\\[0.5ex]%
	  \;  {\rm n} \pi^+ \longrightarrow {\rm n}\;\! \mu^+ \nu_{\upmu}		\\%
		\hspace{5.5em} \myarrow {\rm e}^+ \nu_{\rm e} \bar{\nu}_{\upmu}% 
	\end{cases}  \ .   
\end{align}%  
In addition to pion production, 
   the p$\gamma$ process would lead to 
  Bethe-Heitler leptonic pair-production: 
 \begin{align}
     {\rm p}' + \gamma \rightarrow {\rm p} + l^+ + l^- \ .   
 \end{align} 
Both pp and p$\gamma$ processes 
  can occur while the invariant energy of cosmic-ray collisions with photons and ambient baryons 
  exceeds the required threshold energy. 
 In these processes, a large fraction of neutrons 
  are produced as by-products.  
High-energy hadronic cosmic rays are therefore 
  efficient neutron generators 
  as they propagate and interact with photons and other baryons 
  along their paths. 

%
%======================================================
\subsection{Cosmic ray escape through neutron production} 

\begin{figure}[h]
\centering 
\vspace{-0.5cm}
\includegraphics[scale=0.38]{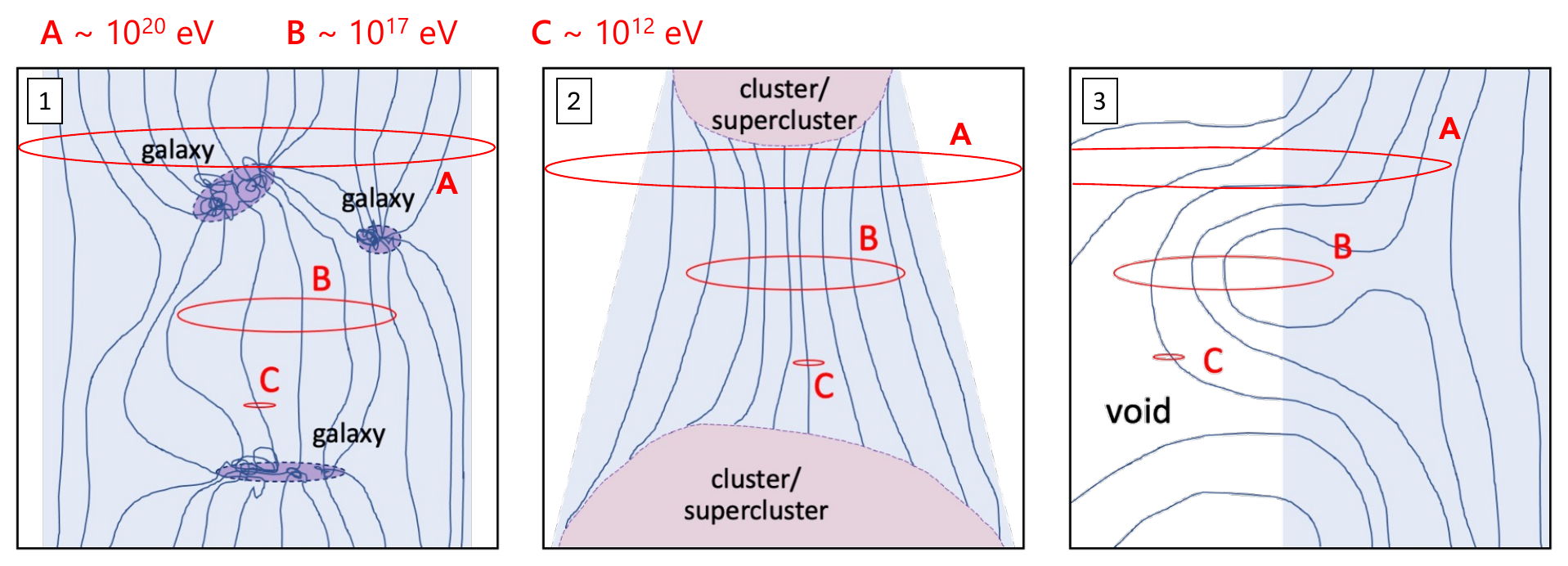}
\caption{Schematic example of three specific situations for the confinement and propagation of energetic particles in cosmic filament environments, and situations that can allow for confinement breaking for particles of different energies. Panel 1 shows the interweaving structure of magnetic field lines connected between a group of galaxies embedded in a filament. Panel 2 shows a filament connecting two clusters/super-clusters, where the filament magnetic field lines have a concave configuration. Panel 3 shows the closed and open magnetic field lines in the interfacing regions of a cosmic filament and a void. The gyration of the charged particles in the magnetic fields may be classified into three regimes represented by the red ellipses, not to scale, labeled as A (cases where particles are not confined), B (particles are weakly confined) and C (particles are strongly confined). The escape of charged particles is possible at high energies, or by confinement breaking in the topologies shown by panels 2 and 3. Figure adapted from \cite{Wu2024Univ}.}
\label{fig:cont_schem}
\end{figure}

UHE cosmic-ray particles, if undeflected,  
travel ballistically in free space   
 at speeds close to that of light. 
Charged cosmic rays are deflected 
  by magnetic fields. 
Astrophysical structures, 
  such as galaxies, 
  are permeated by 
  magnetic fields  
  which, despite being tangled,   
  can have coherent length scales. 
Charged cosmic-rays particles 
   are therefore deflected continuously  
  in magnetized astrophysical structures, 
  making cosmic ray transport 
  effectively diffusive. 
  The confinement of a charged cosmic-ray particle 
  within a domain 
  of a specific magnetic field strength 
  and configuration 
 depends on the energy of the particle 
   (see Figure \ref{fig:cont_schem}). 
Magnetized astrophysical structures therefore 
  practically operate as an energy-dependent sieve 
  for charged cosmic rays. 

Charged cosmic rays can escape 
  from magnetized structures  
  when they acquire energies that are sufficient 
   to exceed the confinement threshold energy.  
This can be achieved, for instance, 
  by in situ acceleration. 
Charged cosmic rays 
  can also break magnetic confinement 
  via cross-field diffusion 
  \citep{Xu2013ApJ}, or 
   by mechanical advection 
  in magnetized flows 
  \citep[e.g.][]{Owen2019MNRAS}. Escape 
  can also be aided by the topology of the field in 
   magnetized structures. Tapered or folded 
   field topologies (e.g. Figure \ref{fig:cont_schem}, panels 2 and 3) 
  could weaken the confinement of particles. 
These confinement-escape scenarios 
   assume 
  that cosmic-ray particles are always charged 
  and particle species are invariant. 
Baryonic cosmic rays  
  consist of uncharged neutrons as well as protons.  Neutrons are not deflected by magnetic fields, so 
the temporary conversion between protons and neutrons   
  via hadronic interactions provides 
  an additional gateway 
  for UHE cosmic rays 
  to escape from magnetized astrophysical structures.  
  
Free neutrons either decay into protons, 
  or they initiate pion-production through either an 
  np process, analogous to the pp interaction (see process \ref{eq:pp_interaction}), or 
   through an n$\gamma$ process: 
\begin{align}%
\label{eq:ng_int}%
 {\rm n} + \gamma \longrightarrow 
		\begin{cases} %
	\; 	{\rm p}\;\! \pi^-  \longrightarrow 
    {\rm p}\;\! \upmu^- \bar{\nu}_{\upmu}		\\ %
		\hspace{5.5em} \myarrow {\rm e}^- {\bar\nu}_{\rm e} {\nu}_{\upmu}% 
    \\  %[0.5ex]%
		\; 	{\rm n}\;\! \pi^0 \longrightarrow {\rm n}\;  2\gamma	  
		\end{cases} \ .   % 
\end{align}%  
These processes occur over 
  a similar length-scale 
  to the corresponding proton pion-production processes. 
Neutrons formed within the dense interstellar medium of galaxies 
  typically experience a hadronic interaction 
  before they escape.
In larger-scale structures, UHE cosmic-ray neutrons 
  can propagate over substantial distances before they decay. 
These energetic neutrons 
  can easily cover distances 
  of a few tens of Mpc 
  across magnetized media in intergalactic space 
  before they decay into protons. 
This allows them to traverse boundaries between structures 
  (e.g. those connecting cosmic filaments and galaxy clusters, 
  or between cosmic filaments and voids) 
  without 
  the aid of cross-field diffusion or mechanical advection.

%
%======================================================
%======================================================
\section{Model, calculations and results} 

%
%======================================================
\subsection{Properties of filaments, clusters and galaxies} 
  
Cosmic filaments are the largest structures in the Universe.  
Their widths are of order a Mpc, 
  and their lengths extend for 10s of Mpc. 
Cosmic filaments are comprised of baryons and dark matter.    
The density of baryons 
  in filaments is about 10-100 times above the cosmic average density in the current epoch \citep{Tanimura2020A&A}. 
These baryons are mostly in the form of hot ionized gas, 
  which intermingles with warm, semi-ionized gas. 
Heating of filament gas is commonly attributed 
   to shocks generated in structure formation and/or feedback processes from lower-order structures (in particular, galaxies).

Observations have indicated 
  that magnetic field strengths in cosmic filaments are of the order 1$-$100~nG 
\citep{Vacca2018Galax, Vernstrom2021MNRAS}. 
Little is known about 
  the structure or topology 
  of these magnetic fields.  
Galaxy clusters and super-clusters   
  are connected by cosmic filaments, 
  and their linear sizes 
  are comparable to filament widths. 
They 
  are relatively young objects 
   in the Universe, and 
the majority of known clusters    
   are found below $z=1$ \citep{Wen2012ApJS}. 
Galaxy clusters have stronger magnetic fields   
   than filaments,   
  especially towards 
  the cluster cores   
  where strengths can reach a few $\mu$G 
  \citep{Vacca2018Galax}. 
Both clusters and filaments  
  harbour galaxies. 
A fraction of these galaxies are isolated systems; 
  some are in pairs or conglomerate into groups. 
Galaxies can be magnetized 
  with field strengths exceeding 
  tens of $\mu$G. 
They acquire their field as they evolve, 
  generally through feedback processes associated with star formation \citep[e.g.][]{Mao2017NatAs}. 
It has also been suggested that 
  their fields could be advected and amplified by inflows 
  from the cosmic web 
  \citep[][]{Ledos2024A&A}. 

\begin{table}
\label{tab:table_of_params}
\caption{Summary of the parameters adopted in our calculations for hadronic pp and p$\gamma$ 
   path lengths for the astrophysical environments shown in 
Fig.~\ref{fig:interaction_lengths}. 
  In all cases, p$\gamma$ interactions with the CMB at $z=0$ are included in our calculations. Additionally, p$\gamma$ interactions interactions with 
    astrophysical radiation fields are considered. These are split into 
  stellar and dust components of the extragalactic background light, which are modeled 
 as modified black-bodies, 
 with characteristic 
 temperatures of 7,100 K (starlight) and 62 K (dust) with the indicated energy densities, $U_{\rm rad}$. 
 See \cite{Wu2024Univ} for an explanation 
  of these parameter choices.}
 \small
\begin{center}
\begin{tabular}{|l|cc|c|c|c|}
\hline
\multirow[m]{2}{*}{\textbf{Structure}} & \multicolumn{2}{|c|}{\textbf{$U_{\rm rad}$ [eV cm$^{-3}$]}} & \textbf{Gas density} & \textbf{Magnetic} & \textbf{Size} \\
 & \textit{Starlight} & \textit{Dust}  & \textbf{[g cm$^{-3}$]} &  \textbf{field [G]} & \textbf{[Mpc]} \\
\hline
Filament              & 3.7          & 5.2 & $4.0\times 10^{-29}$ & $10^{-9}$ & 2.0 \\
\hline
Cluster             & 0.21         & 0.28 & $1.1\times 10^{-27}$ & $10^{-6}$ & 1.9 \\
\hline
Galaxy               & \multirow[m]{2}{*}{670}          & \multirow[m]{2}{*}{310} & \multirow[m]{2}{*}{$1.7\times10^{-20}$} &  \multirow[m]{2}{*}{$10^{-5}$} & \multirow[m]{2}{*}{0.0010}\\
(starburst) & & & & & \\
\hline
\end{tabular}
\end{center}
\vspace{-0.7cm}
\end{table}

In our calculations,  
we have adopted 
  a simple hierarchical prescription 
  of the embedded structures in filaments 
  to assess 
  the survival, propagation and escape 
  of UHE protons and neutrons. 
The variables - 
  energy density of their radiation fields, 
  gas density, sizes of objects, 
  and magnetic field strengths --  
  are summarized in Table \ref{tab:table_of_params} 

%
%======================================================
\subsection{Results and implications} 

\begin{figure}[h]
\centering
\includegraphics[scale=0.43]{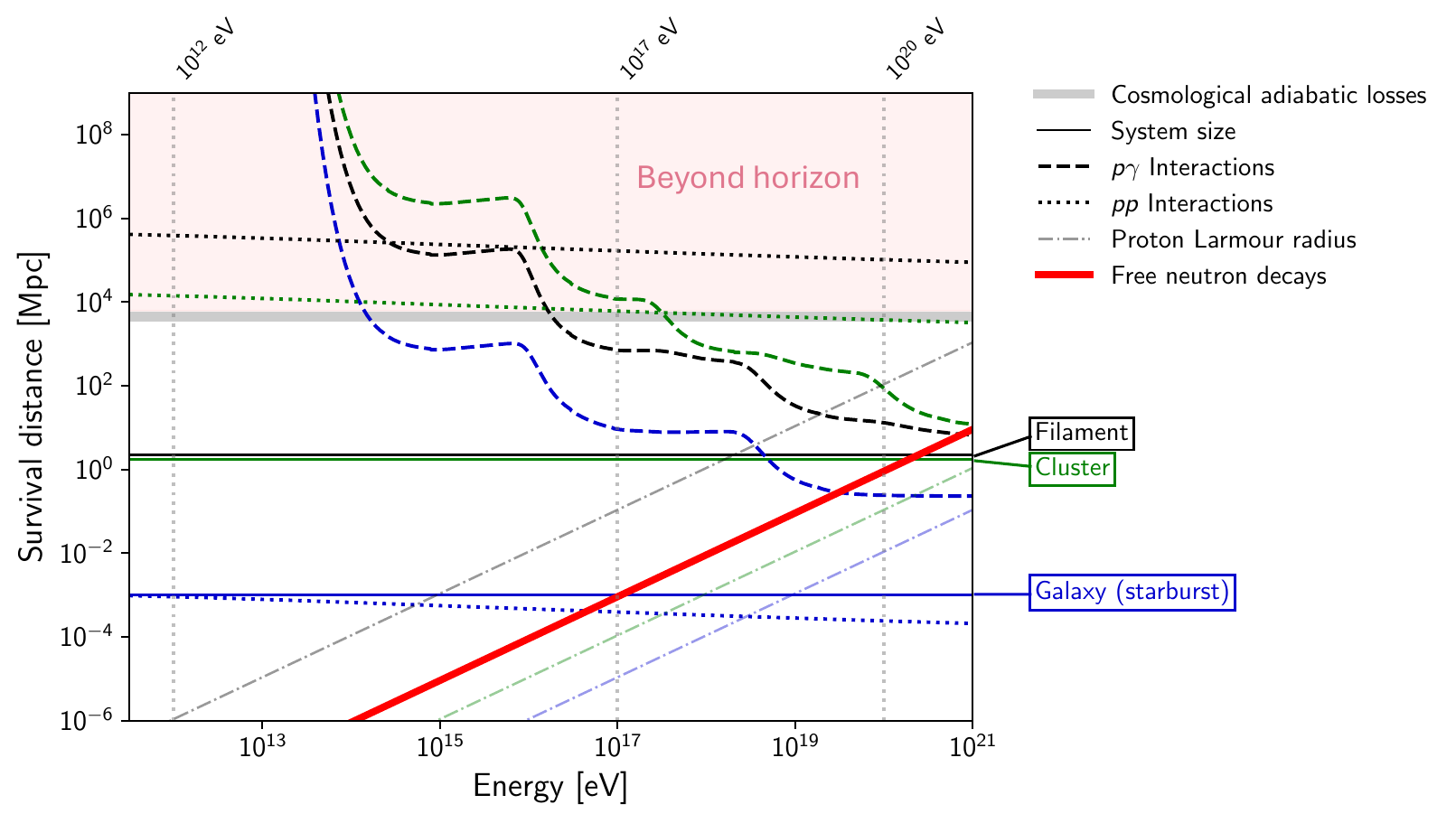}
\vspace*{-0.2cm}
\caption{Characteristic propagation distances of hadronic cosmic-ray protons and neutrons subject to pp and p$\gamma$ interactions and decays, compared to the size of their magnetized host structures (blue lines correspond to path lengths in galaxy environments, green for clusters and black for cosmic filaments, as labeled). For reference, the free neutron decay length is indicated in red, and the characteristic gyro-radius for cosmic-ray protons is shown for each of the structures. The distance to the event horizon of the Universe at the current epoch is roughly the same as the adiabatic loss length scale of protons. Interaction lengths above this scale (indicated by the pink shaded region) are not of astrophysical consequence, but are shown for completeness. Vertical lines indicate the energies of the particle trajectories A, B and C shown in Fig. \ref{fig:cont_schem}.}
\label{fig:interaction_lengths}
\end{figure}

 In Figure \ref{fig:interaction_lengths}, we show 
  the survival distances 
  of cosmic-ray protons 
  subject to pp and p$\gamma$ processes in 
   filaments, clusters and galaxies. 
  We also show the cosmic-ray proton gyro-radius 
  in magnetic field strengths relevant 
  to these environments. 
This sets the reference scale 
  for cosmic-ray containment.  
Particle escape below the containment limit is still possible but depends on diffusion and advection, which are most efficient for low-energy cosmic rays trapped by magnetic fields.
The free neutron survival distance indicates the structures neutrons could traverse prior to decaying. Magnetic confinement can be bypassed over these scales if cosmic-ray protons are converted to neutrons via hadronic interactions within each of the  structures. 

Our calculations show that  
galaxies operate as proton calorimeters 
up to UHE energies. The short pp interaction length-scale means they efficiently convert protons into neutrons at all energies. Up to $\sim 10^{17}$ eV, neutrons do not propagate far before decaying back into protons.  
Despite this, a 
relatively high neutron fraction could be sustained 
in the dense interstellar medium of galaxies 
as baryons continue to undergo hadronic interactions. At higher energies, the neutron decay length becomes comparable to the size of a galaxy, and cosmic ray confinement could be broken after just a single hadronic interaction. Both of these 
factors could allow a moderate fraction of cosmic rays to escape galaxies, especially above $\sim 10^{17}$ eV. 

Based on the arguments of continuity of magnetic field lines  
  and magnetic flux conservation, 
  \cite{Wu2024Univ} pointed out that  
  there would be a magnetic barrier 
  at cluster - filament interfaces.  
This barrier only allows 
  charged cosmic rays to traverse the boundary 
  with the aid of cross-field diffusion or bulk advective flows. 
Figure \ref{fig:cont_schem} illustrates that, above $\sim 10^{19}$ eV, a non-negligible fraction of cosmic rays can undergo 
 a photo-hadronic interaction as they propagate through filaments. The subsequent neutrons that form could easily pass the interface into clusters, and their subsequent decay to protons would lead to confinement by the more strongly magnetized intra-cluster medium. Cluster boundaries can therefore be considered as permeable to UHE cosmic rays, but operate as an energy-dependent sieve preventing the passage of cosmic rays below $\sim 10^{19}$ eV, including charged cosmic rays confined and channeled along filaments. Similar to the case of clusters~\citep[see][]{Fang2016ApJ, Hussain2023NatCo}, this scenario could be tested through future measurements of $\gamma$-rays and neutrinos (being by-products of neutron production) from filaments using instruments with sufficient sensitivity and appropriate data analysis techniques.

Cosmic-ray protons in clusters are expected to be confined at all energies. However, above $\sim 10^{20}$ eV, they will undergo photo-hadronic interactions over timescales of shorter than a Gyr. The resulting neutrons from these interactions are sufficiently energetic to easily break confinement. Given the long diffusion timescales and the absence of alternative cooling or loss mechanisms, this can form a major channel for UHE cosmic ray escape from galaxy clusters/super-clusters. 

\section{Remarks and Conclusions}

\begin{figure}[h]
\centering
\begin{minipage}{0.6\textwidth}
    \centering
    \includegraphics[scale=0.37]{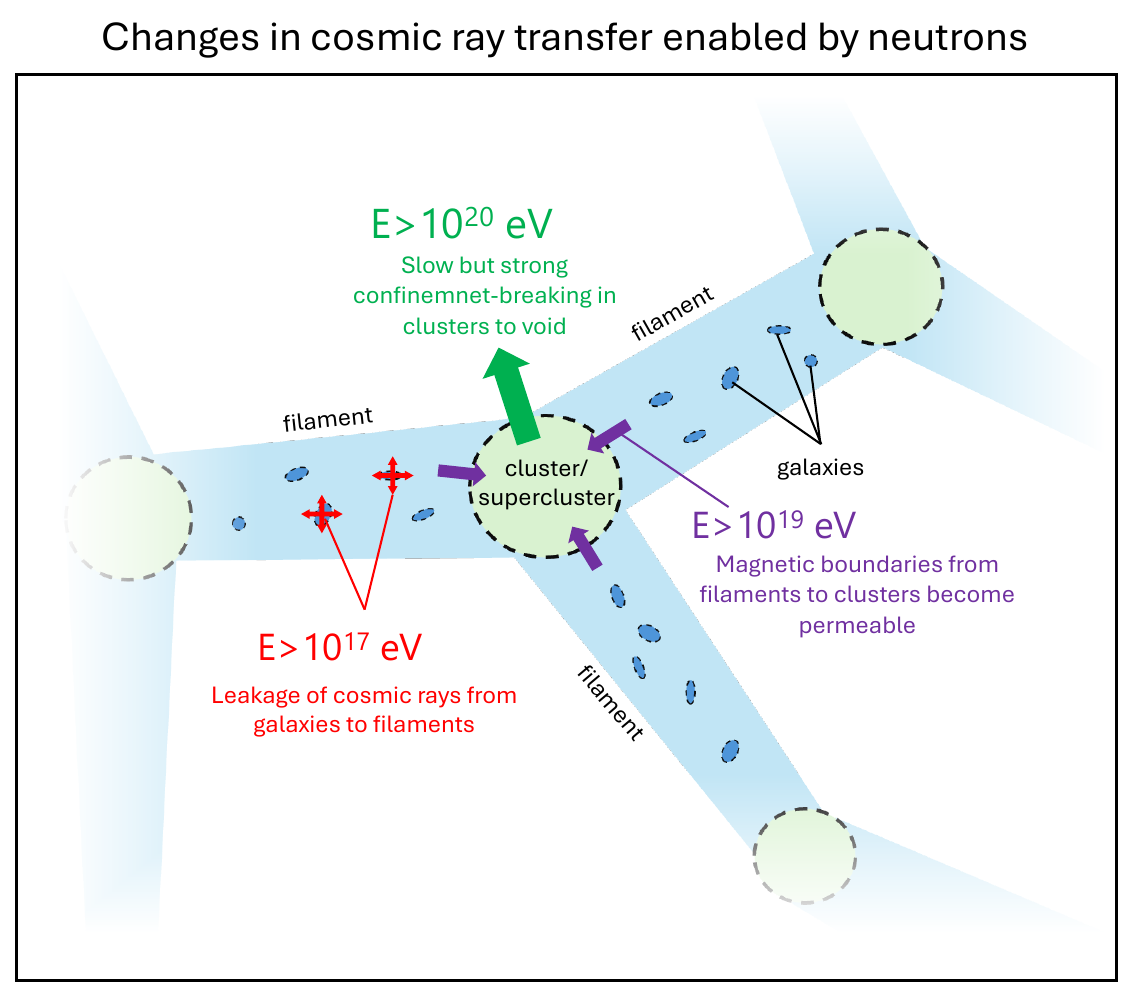}
\end{minipage}%
\hfill
\begin{minipage}{0.39\textwidth}
    \caption{Schematic to show the main ways cosmic rays can pass between filaments and their embedded structures that are enabled by neutron production. At energies substantially below $10^{17}$ eV, changes can only arise if hadronic interactions produce neutrons near structural boundaries or if they are continually formed in regions where interaction rates are high. At higher energies, neutron path lengths become comparable to the sizes of their containing structures with more consequential impacts on cosmic ray confinement.}
    \label{fig:escape_channels}
\end{minipage} 
\vspace*{-0.35cm}
\end{figure}

Figure \ref{fig:escape_channels} illustrates the main escape channels that can be facilitated by neutron production via hadronic cosmic ray interactions in filaments, clusters and galaxies. We conclude that neutrons have the largest effects on cosmic ray confinement at the highest of energies. They can enable the escape of cosmic rays from clusters above $10^{20}$ eV, and allow boundaries between lower order structures to be permeated. 
Galaxy clusters trap and attenuate UHE cosmic rays \citep{Condorelli2023ApJ}, yet our results suggest that these effects can be overcome at the highest of energies, where a gradual leaking of their cosmic ray population can be sustained by photo-hadronic interactions.  Clusters therefore remain viable as a possible origin for some of the highest energy cosmic rays detected on Earth.

\vspace{-0.3cm}

\vspace{0.0cm}
\footnotesize
\noindent
\textbf{Acknowledgements:} 
E.R.O. is an overseas researcher supported by a Postdoctoral Fellowship of the Japan Society for the Promotion of Science (JSPS KAKENHI Grant JP22F22327), and also acknowledges support from the RIKEN Special Postdoctoral Researcher Program for junior scientists. 
Y.I. is supported by an NAOJ ALMA Scientific Research Grant 2021-17A, JSPS KAKENHI Grants JP18H05458, JP19K14772, and JP22K18277, and the World Premier International Research Center Initiative (WPI), MEXT, Japan. 
Q.H. is supported by a UCL Overseas Research Scholarship and a UK Science and Technology Facilities Council Research Studentship. We thank the anonymous referee for their constructive comments. 
\vspace{-0.3cm}

%%%%%%%%%%%%%%%%%%%%%%%%%%%%%%%%%%%%

%%%%%%%%%%%%%%%%%%%%%%%%%%%%%%%%%%%%%
%References produced by BibTeX
\bibliographystyle{apalike}
\bibliography{references}

\begin{thebibliography}{}

\bibitem[{Bell}, 1978]{Bell1978MNRAS}
{Bell}, A.~R. (1978).
\newblock {The acceleration of cosmic rays in shock fronts - I.}
\newblock {\em \mnras}, 182:147--156.

\bibitem[{Condorelli} et~al., 2023]{Condorelli2023ApJ}
{Condorelli}, A., {Biteau}, J., and {Adam}, R. (2023).
\newblock {Impact of Galaxy Clusters on the Propagation of Ultrahigh-energy
  Cosmic Rays}.
\newblock {\em \apj}, 957(2):80.

\bibitem[{Fang} and {Olinto}, 2016]{Fang2016ApJ}
{Fang}, K. and {Olinto}, A.~V. (2016).
\newblock {High-energy Neutrinos from Sources in Clusters of Galaxies}.
\newblock {\em \apj}, 828(1):37.

\bibitem[{Fermi}, 1949]{Fermi1949PhRv}
{Fermi}, E. (1949).
\newblock {On the Origin of the Cosmic Radiation}.
\newblock {\em Physical Review}, 75(8):1169--1174.

\bibitem[{Hussain} et~al., 2023]{Hussain2023NatCo}
{Hussain}, S., {Alves Batista}, R., {de Gouveia Dal Pino}, E.~M., and {Dolag},
  K. (2023).
\newblock {The diffuse gamma-ray flux from clusters of galaxies}.
\newblock {\em Nature Communications}, 14:2486.

\bibitem[{Ledos} et~al., 2024]{Ledos2024A&A}
{Ledos}, N., {Ntormousi}, E., {Takasao}, S., and {Nagamine}, K. (2024).
\newblock {Magnetising galaxies with cold inflows}.
\newblock {\em \aap}, 691:A280.

\bibitem[{Mao} et~al., 2017]{Mao2017NatAs}
{Mao}, S.~A., {Carilli}, C., {Gaensler}, B.~M., {Wucknitz}, O., {Keeton}, C.,
  {Basu}, A., {Beck}, R., {Kronberg}, P.~P., and {Zweibel}, E. (2017).
\newblock {Detection of microgauss coherent magnetic fields in a galaxy five
  billion years ago}.
\newblock {\em Nature Astronomy}, 1:621--626.

\bibitem[{M{\"u}cke} et~al., 1999]{Mucke1999PASA}
{M{\"u}cke}, A., {Rachen}, J.~P., {Engel}, R., {Protheroe}, R.~J., and
  {Stanev}, T. (1999).
\newblock {Photohadronic Processes in Astrophysical Environments}.
\newblock {\em \pasa}, 16(2):160--166.

\bibitem[{Owen} et~al., 2019]{Owen2019MNRAS}
{Owen}, E.~R., {Jin}, X., {Wu}, K., and {Chan}, S. (2019).
\newblock {Hadronic interactions of energetic charged particles in
  protogalactic outflow environments and implications for the early evolution
  of galaxies}.
\newblock {\em \mnras}, 484(2):1645--1671.

\bibitem[{Particle Data Group}, 2020]{PDG2020PTEP}
{Particle Data Group} (2020).
\newblock {Review of Particle Physics}.
\newblock {\em Progress of Theoretical and Experimental Physics},
  2020(8):083C01.

\bibitem[{Sato}, 2015]{Sato2015PLoSO}
{Sato}, T. (2015).
\newblock {Analytical Model for Estimating Terrestrial Cosmic Ray Fluxes Nearly
  Anytime and Anywhere in the World: Extension of PARMA/EXPACS}.
\newblock {\em PLoS ONE}, 10(12):e0144679.

\bibitem[{Schimassek} et~al., 2024]{Schimassek2024arXiv}
{Schimassek}, M., {Engel}, R., {Ferrari}, A., {Roth}, M., {Schmidt}, D., and
  {Veberi{\v{c}}}, D. (2024).
\newblock {Neutron Production in Simulations of Extensive Air Showers}.
\newblock {\em arXiv e-prints}, page arXiv:2406.11702.

\bibitem[{Tanimura} et~al., 2020]{Tanimura2020A&A}
{Tanimura}, H., {Aghanim}, N., {Kolodzig}, A., {Douspis}, M., and {Malavasi},
  N. (2020).
\newblock {First detection of stacked X-ray emission from cosmic web
  filaments}.
\newblock {\em \aap}, 643:L2.

\bibitem[{Vacca} et~al., 2018]{Vacca2018Galax}
{Vacca}, V., {Murgia}, M., {Govoni}, F., {En{\ss}lin}, T., {Oppermann}, N.,
  {Feretti}, L., {Giovannini}, G., and {Loi}, F. (2018).
\newblock {Magnetic Fields in Galaxy Clusters and in the Large-Scale Structure
  of the Universe}.
\newblock {\em Galaxies}, 6(4):142.

\bibitem[{Vernstrom} et~al., 2021]{Vernstrom2021MNRAS}
{Vernstrom}, T., {Heald}, G., {Vazza}, F., {Galvin}, T.~J., {West}, J.~L.,
  {Locatelli}, N., {Fornengo}, N., and {Pinetti}, E. (2021).
\newblock {Discovery of magnetic fields along stacked cosmic filaments as
  revealed by radio and X-ray emission}.
\newblock {\em \mnras}, 505(3):4178--4196.

\bibitem[{Wen} et~al., 2012]{Wen2012ApJS}
{Wen}, Z.~L., {Han}, J.~L., and {Liu}, F.~S. (2012).
\newblock {A Catalog of 132,684 Clusters of Galaxies Identified from Sloan
  Digital Sky Survey III}.
\newblock {\em \apjs}, 199(2):34.

\bibitem[{Wu} et~al., 2024]{Wu2024Univ}
{Wu}, K., {Owen}, E.~R., {Han}, Q., {Inoue}, Y., and {Luo}, L. (2024).
\newblock {Energetic Particles and High-Energy Processes in Cosmological
  Filaments and Their Astronomical Implications}.
\newblock {\em Universe}, 10(7):287.

\bibitem[{Xu} and {Yan}, 2013]{Xu2013ApJ}
{Xu}, S. and {Yan}, H. (2013).
\newblock {Cosmic-Ray Parallel and Perpendicular Transport in Turbulent
  Magnetic Fields}.
\newblock {\em \apj}, 779(2):140.

\bibitem[{Ziegler}, 1998]{Ziegler1998IBMJ}
{Ziegler}, J.~F. (1998).
\newblock {Terrestrial cosmic ray intensities.}
\newblock {\em IBM Journal of Research and Development}, 42(1):117--139.

\end{thebibliography}
%%%%%%%%%%%%%%%%%%%%%%%%%%%%%%%%

\end{document}